\documentclass[a4paper]{jpconf}
\usepackage{graphicx}
\usepackage{amsmath}
\begin{document}
\title{Effective approach to top-quark decay and FCNC processes at NLO accuracy}

\author{Cen Zhang}

\address{
Centre for Cosmology, Particle Physics and Phenomenology,
Universit\'e catholique de Louvain, 2 Chemin du Cyclotron,
B-1348 Louvain-la-Neuve, Belgium
}

\ead{cen.zhang@uclouvain.be}

\begin{abstract}
  The top quark is expected to be a probe to new physics beyond the standard
  model.  Thanks to the large number of top quarks produced at the Tevatron and
  the LHC, various properties of the top quark can now be measured
  accurately.  An effective field theory allows us to study the new physics
  effects in a model-independent way, and to this end accurate theoretical
  predictions are required.  In this talk we will discuss some recent results on
  top-quark decay processes as well as flavor-changing processes, based on the
  effective field theory approach.
\end{abstract}

\section{Introduction}

Current strategies to search for new physics beyond the standard model (SM) can
be broadly divided into two categories.  In the first category we look for new
resonant states.  In the second category, new states are assumed to be heavy,
and we look for their indirect effects in the interactions of known particles.

The top quark has been a natural probe to new physics in the first category,
due to its large mass and strong coupling to the electroweak symmetry breaking
sector.  Searches for resonant states through decay processes involving the top
quarks have been performed both at the Tevatron and at the LHC.  Examples
include $t\bar t$ resonance searches, top partner production, and so on.
Unfortunately, until now no new states have been discovered, and exclusion
limits have been placed, up to around several TeV scale.

On the other hand, the top quark physics has entered a precision era, thanks to
the large number of top quarks produced at the Tevatron and LHC.  Various
properties of the top quark have been already measured with high precision, and
the upcoming LHC Run-II will continue to increase the precision level.
From the theory side, accurate SM predictions are also available.  As a result
the focus of top quark physics is now moving toward the second category, i.e.~to
measure accurately the known interactions and the rare processes of SM
particles.  Examples are measurements on $W$-helicity fractions in top-quark
decay, and searches for processes involving flavor-changing neutral current
(FCNC) of the top quark.  These are the main topics of this talk.

\section{The effective approach}

When looking for deviations from the SM interactions, the standard approach is to
utilize the effective field theory (EFT) framework, in which deviations from the
SM are parameterized by including higher dimensional operators.  The approach is
valid when the new physics scale, $\Lambda$, is higher than the scale of the
process.  Assuming the full SM gauge symmetries, the leading new operators
involving a top quark arise at dimension six. At this level
the Lagrangian can be written as
\begin{equation}
  \mathcal{L}_{EFT}=\mathcal{L}_{SM}+\sum_i\frac{C_i}{\Lambda^2}O_i+\mathrm{h.c.}
\end{equation}
A calculation in the EFT framework consists of three steps:
\begin{enumerate}
  \item Matching calculation.  The EFT is matched to the full theory which
    lives at the high energy scale $\Lambda$.  This determines the operator
    coefficients $C_i$, defined at the matching scale.
  \item Renormalization group (RG) evolution.  The coefficients will be evolved
    down to a lower scale $E$ where the actual process takes place. This
    essentially resums the $\log(\Lambda/E)$ terms.
  \item Matrix element calculation, to be performed at scale $E$.
\end{enumerate}

The procedures are familiar from the RG-improved perturbation theory which
plays an important role in flavor physics.  Indeed there is a similarity
between top quark physics and flavor physics.  In flavor physics, the
full theory is the SM, possibly with some new physics.  The matching is
performed at scale $\Lambda=m_W$, where the heavy SM particles are integrated
out.  The theory is then evolved down to lower scales where the matrix element
can be computed either perturbatively or non-perturbatively.  In the top quark
physics, the full theory is the SM plus new physics, which is to be matched to
the EFT at scale $\Lambda=\Lambda_{NP}$, where the new physics lives.  The
procedure integrates out heavy particles in the new physics, but the heavy
particles of the SM remain in the effective theory, leading to 59 dimension-six
operators in total for one generation of fermions \cite{Grzadkowski:2010es}.
We then select the ones that are relevant for top-quark processes and evolve
them down to a scale of $\sim m_t$.  Cross sections can then be computed in
terms of these operator coefficients.

Recently the RG equations for all dimension-six operators have been derived at
the one-loop level \cite{Jenkins:2013zja,Jenkins:2013wua,Alonso:2013hga},
completing the second step in this picture.  There are also new developments in
matrix element calculations.  In particular, the top-quark decay and several FCNC
processes have been computed at NLO in QCD, and the automation of the FCNC
processes in the MG5\_aMC@NLO framework \cite{Alwall:2014hca} is now in progress.
These will be discussed in the remainder of this talk. 

\section{Top-quark decay and $W$-helicity fractions}

In this section we will discuss the main decay channel of the top quark,
$t\to bW$, focusing on
the $W$-helicity fractions. For SM couplings and unpolarized top-quark
production, the helicity fractions are approximately 70\% longitudinal and 30\%
left-handed.  Accurate predictions at NNLO in QCD and NLO in electroweak are
available.

In an EFT, dimension-six operators will modify the helicity fractions.  First
of all, there are operators that directly modify the standard $Wtb$ vertex
function.  They are
\begin{equation}
  \begin{array}{ll}
    O_{\varphi Q}^{(3)}
    =i\frac{1}{2}y_t^2 \left(\varphi^\dagger\overleftrightarrow{D}^I_\mu\varphi\right)
    (\bar{Q}\gamma^\mu\tau^I Q)
    &
    O_{tW}=y_tg_W(\bar{Q}\sigma^{\mu\nu}\tau^It)\tilde{\varphi}W_{\mu\nu}^I
    \\
    O_{\varphi\varphi}=iy_t^2\left(\tilde{\varphi}^\dagger
    D_\mu\varphi\right)(\bar{t}\gamma^\mu b)
    &
    O_{bW}=y_tg_W(\bar{Q}\sigma^{\mu\nu}\tau^Ib)\varphi W_{\mu\nu}^I
  \end{array}
\end{equation}
where $Q$ is the left-handed ($t$, $b$) doublet. The QCD corrections to these
operators have been computed in Ref.~\cite{Drobnak:2010ej}.  Deviations from
the SM predictions are expected to be dominated by contribution from $O_{tW}$.

A second possibility comes from the color-dipole operator,
\begin{equation}
  O_{tG}=y_tg_s(\bar{Q}\sigma^{\mu\nu}T^At)\tilde{\varphi}G_{\mu\nu}^A
\end{equation}
which only contributes at NLO in QCD.  This has been computed in
Ref.~\cite{Zhang:2014rja}.
There is a mixing effect from operator
$O_{tG}$ to $O_{tW}$, given by $\gamma_{C_{tW},C_{tG}}=2\alpha_s/3\pi$.

Finally, four-fermion operators such as 
$O_{lQ}^{(3)}=\left(
\bar{l}\gamma_\mu\tau^Il \right) \left( \bar{Q}\gamma^\mu\tau^IQ \right)$
should also be considered.  These operators arise from for example a heavy $W'$
particle exchange, and can directly contribute to the $bl\nu$ final state.  Their
contributions to the total decay rate are small.  Nevertheless we can
hope to set some bounds using differential decay rate, since they will affect
not only the measured $F_+$ value (which would be tiny in the SM), but also the
$m_{l\nu}$ spectrum. 

The helicity fractions are measured both at the Tevatron and at the LHC.  Using
the results from Ref.~\cite{Chatrchyan:2013jna}, one can already set some
constraints.  For illustration, assuming other operators vanish, the limit
on $O_{tW}$ is
\begin{equation}
  C_{tW}=-0.11\pm1.13\,(\Lambda/\rm{TeV})^2
\end{equation}
at 95\% CL.
However one should keep in mind that considering only one operator at a time is
not natural, in particular due to the operator mixing effects. In principle,
four-fermion operators should also be considered, however current strategy to
top-quark reconstruction requires that $m_{l\nu}$ is equal to the $W$ mass.
This condition does not apply when four-fermion operators are present, so the
current limits on the helicity fractions cannot be used to constrain
four-fermion operators.

\section{Top-quark decay via FCNC}

Flavor-changing decay modes such as $t\to qX$ are suppressed by the
Glashow-Iliopoulos-Maiani mechanism, and the SM predictions for the branching
ratios are of order $10^{-12}-10^{-15}$.  Any evidence for these decay modes
would be a definite signal for new physics.  In the EFT framework the decay
rates for $t\to ug$, $t\to u\gamma$, $t\to uh$ and $t\to ull$ are all computed
at NLO in QCD \cite{Drobnak:2010wh,Zhang:2010bm,Zhang:2014rja}.  Relevant
operators can be found in \ref{sec:app}.
On the experimental side,
branching ratios for $u\gamma$, $ull$ and $uh$ final states have been measured
both at the Tevatron and at the LHC.  Single top production processes
via FCNC couplings
are also studied, via $e^+e^-\to t\bar q$ at LEP2, $e^-q\to e^-t$ at
HERA, and $qg\to t(j)$ at the Tevatron and the LHC.  Limits on FCNC couplings
obtained from production processes are often converted to limits on branching
ratios.

In general, a single process can depend on many operators, and so one
measurement can only set bounds on a specific linear combination of operator
coefficients.  Once we combine all available measurements, a global analysis
can be performed to determine the allowed region for all operator coefficients.
However, limits obtained by experimental collaborations almost always assume
only one FCNC interaction is present at a time.  In addition, four-fermion
operators are always neglected in both production and decay processes.  For
these two reasons it is difficult to perform a real global fit based on the EFT
framework.

To illustrate the feasibility of such a global approach, here we present an
``approximate'' global fit for the top quark FCNC sector.  We neglect all
four-fermion operators, and only consider $t\to qZ$ \cite{Chatrchyan:2013nwa}, $t\to
qh$ \cite{CMS:2014qxa} \footnote{This limit applies for $t\to ch$, but limit
  for $t\to uh$ is expected to be better \cite{Greljo:2014dka}.
  A slightly weaker bound \cite{Aad:2014dya} applies for both decay modes.
} and $pp\to t$
\cite{TheATLAScollaboration:2013vha} via FCNC couplings involving an up
quark.  Moreover we interpret the limit from $pp\to t\gamma$
\cite{CMS:2014hwa} as a direct bound on the branching ratio of $t\to
q\gamma$, which is much tighter than the current bound measured at the
Tevatron.  These four measurements suffice to constrain all two-quark FCNC
operators.  Unfortunately, the published information is not enough for a
consistent combination of all four measurements at 95\% CL, therefore we
only require that all constraints at 95\% CL are independently satisfied.

\begin{figure}[t]
  \begin{minipage}{.55\linewidth}
    \begin{center}
      \includegraphics[width=.9\linewidth]{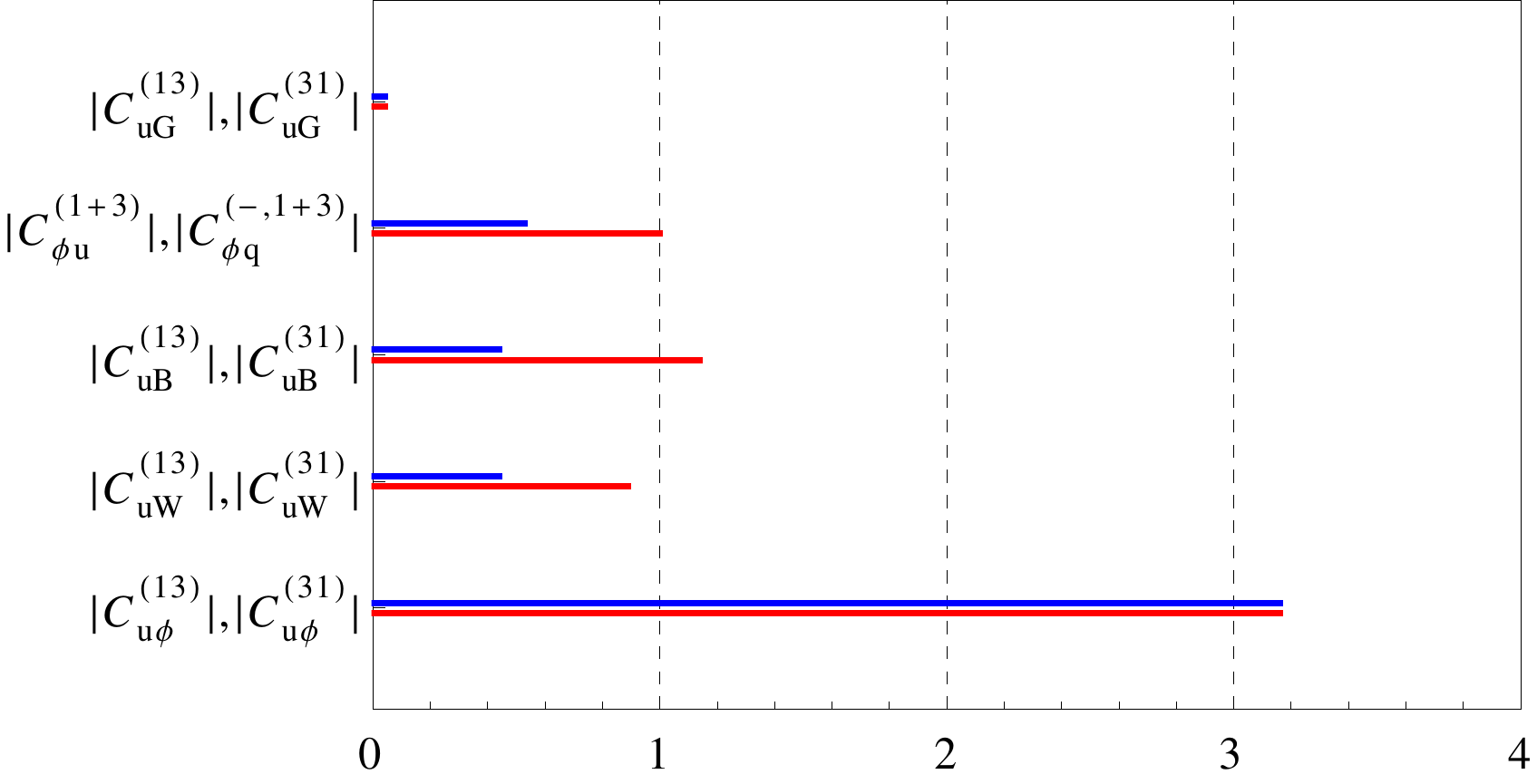}
    \end{center}
    \caption{\label{fig:limits}Limits on individual operator coefficients. Blue lines
  indicate limits obtained by setting other coefficients to zero. Red lines are
obtained by allowing other coefficients to float. $\Lambda=1$ TeV is assumed.}
  \end{minipage}
  \hfill
  \begin{minipage}{.4\linewidth}
    \begin{center}
      \includegraphics[width=.99\linewidth]{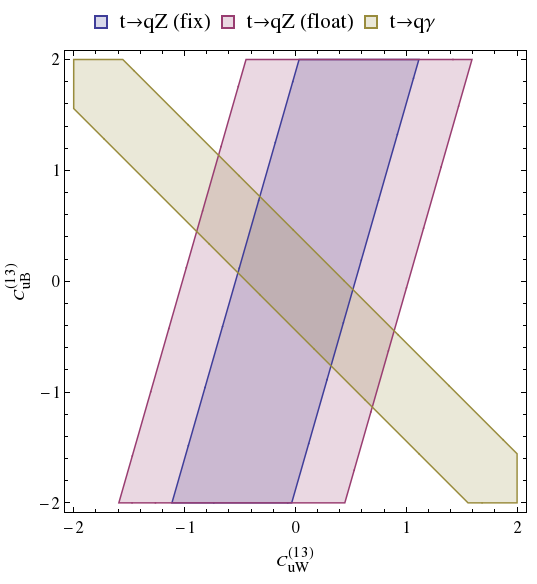}
    \end{center}
    \caption{\label{fig:ubuw}Allowed region in $C_{uW}^{(13)}-C_{uB}^{(13)}$
  space, from $t\to uZ$ and $pp\to t\gamma$, by fixing other coefficients to
zero and by allowing them to float.  $\Lambda=1$ TeV.}
  \end{minipage}
\end{figure}

The result of the fit is shown in Figure~\ref{fig:limits}. 
The blue lines are obtained by setting other coefficients to
zero, while the red lines are obtained by allowing other coefficients to float.
For some operators the blue and red lines are different, indicating that a
correlation is present between the operators.  This effect between
$O_{uB}$ and $O_{uW}$ can be observed in Figure~\ref{fig:ubuw}.

The situation can be more complicated if four-fermion operators are included.
These operators may contribute to the decay channel $t\to qll$, which is the
actual final state of the $t\to qZ$ measurements.  Consider for example the
following operator
\begin{equation}
  O_{lequ}^{(3,13)}=\left( \bar l \sigma_{\mu\nu}e \right)\varepsilon
  \left( \bar q\sigma^{\mu\nu}t \right)\ ,
\end{equation}
its contribution is comparable with typical contributions from two-fermion
operators.  Therefore constraining four-fermion operator coefficients from $t\to
qZ$ is possible.

\begin{figure}[t]
    \includegraphics[width=.5\linewidth]{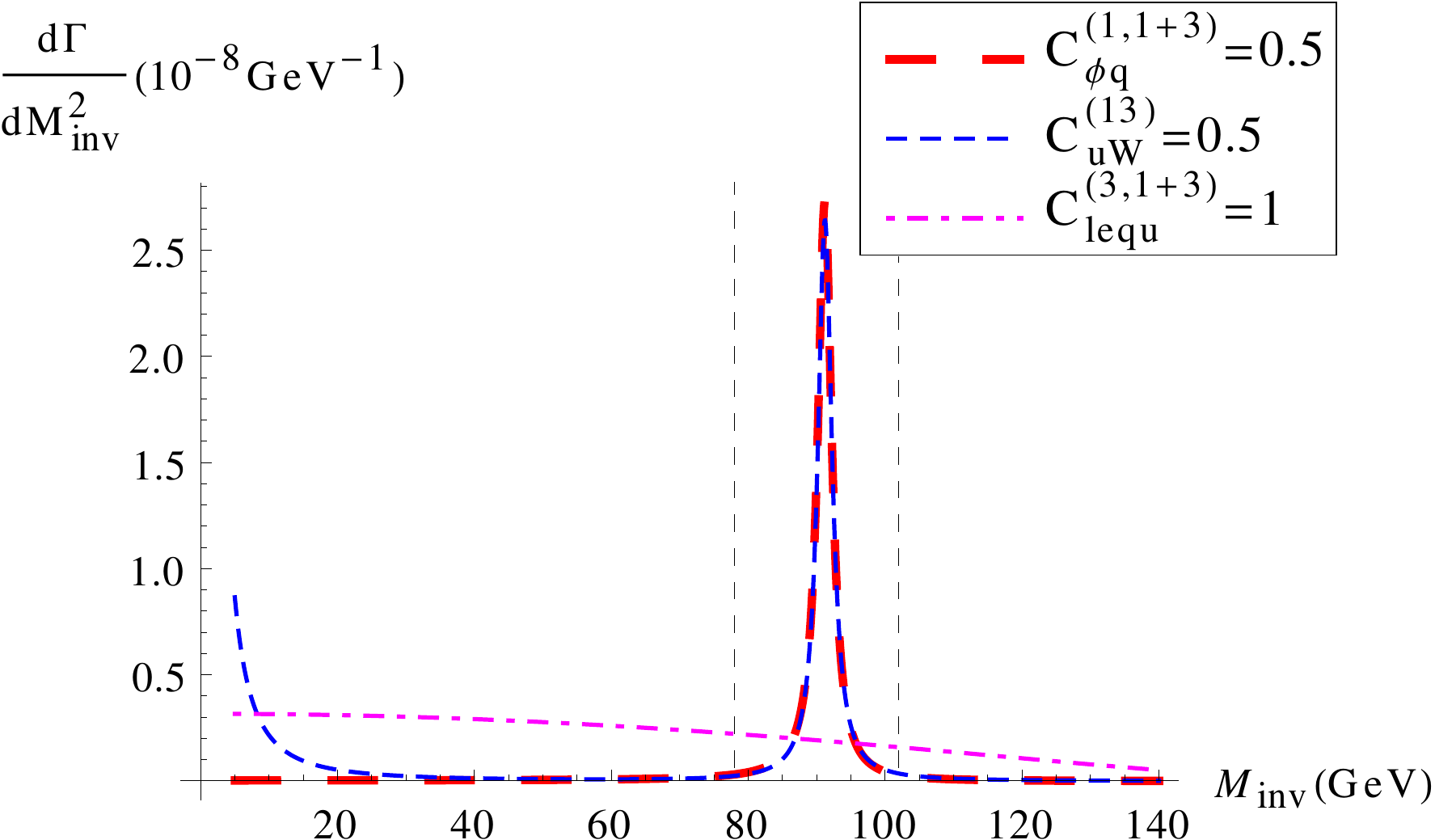}
    \hfill
  \begin{minipage}[b]{.4\linewidth}
    \caption{\label{fig:4fdecay}Invariant mass distribution of lepton pairs from
    FCNC top-quark decay.  Contributions from two-fermion operators, $O_{\varphi
    q}^{(1,1+3)}$ and $O_{uW}^{(13)}$, and four-fermion operator
    $O_{lequ}^{(3,1+3)}$ are compared. $\Lambda$=1 TeV.}
    \end{minipage}
\end{figure}
The main difference between two- and four-fermion operators is reflected by the
invariant mass distribution of the lepton pairs.  Two-fermion operators involve
a $Z$ boson and show a peak near the $Z$ mass, while four-fermion operators
give a more flat spectrum.  We illustrate this difference in
Figure~\ref{fig:4fdecay}, where the four-fermion operator $O_{lequ}^{(3,1+3)}$
and two additional two-fermion operators $O_{\varphi q}^{(1,1+3)}$ and
$O_{uW}^{(13)}$ are included.  Current strategies to searching for FCNC
top-quark decay impose some on-shell cuts on the invariant mass of the two
leptons, e.g.~$m_{ll}\in[78,102]$ GeV, to select events with two
leptons coming from an on-shell $Z$ boson.  These are displayed in
Figure~\ref{fig:4fdecay} as two dashed straight lines.  After applying these
cuts, the contributions from the three operators are:
\begin{equation}
  \Gamma_{\rm on-shell}=
  \left(7.0|{C_{uW}^{(13)}}|^2
  +7.3|{C_{\varphi q}^{(1,1+3)}}|^2
  +0.8|{C_{lequ}^{(3,13)}}|^2\right)
  \times 10^{-5}\ \mathrm{GeV}\,.
  \label{eq:onshellrate}
\end{equation}
This implies that even with the on-shell cuts, the four-fermion operator
is still not negligible.  In fact, the above results will lead to limit on
$C_{lequ}^{(3,13)}$ that is only 3 times larger than those of
the two-fermion operator coefficients.

On the other hand, looking at the off-shell regions of the spectrum
will give us more information.  The decay rate in 
$m_{ll}\in[15,78]\cup[102,\infty]$ GeV
is:
\begin{equation}
  \Gamma_{\rm off-shell}=
  \left(0.6|
  {C_{uW}^{(13)}}|^2+0.4|
  {C_{\varphi q}^{(1,1+3)}}|^2
  +2.7|{C_{lequ}^{(3,13)}}|^2\right)
  \times 10^{-5}\ \mathrm{GeV}\,.
    \label{eq:offshellrate}
\end{equation}
We can see that two-fermion contributions are suppressed while the four-fermion
one is enhanced, so by looking at both regions we can constrain two- and
four-fermion operators separately.  Due to less Drell-Yan background in the off-shell
region, one might even get a better sensitivity on four-fermion operators.

\section{Single Top Production via FCNC}

We now turn to single top production via top-quark FCNC.  Recently
single top production associated with a photon and with a $Z$-boson have been
searched for at the LHC \cite{CMS:2014hwa,CMS:2013nea}.  Limits are obtained on
the FCNC couplings $tqg$, $tq\gamma$ and $tqZ$.  Single top production
associated with a Higgs boson has also been proposed as a probe of
the $tqg$ and $tqh$ couplings.  These analyses are important consistency checks
for the FCNC searches already performed in top-decay processes.  In a global
fit based on an EFT approach, they can be combined together with the decay
processes, to improve the limits on various FCNC couplings.

\begin{figure}[tb]
\begin{minipage}{0.46\linewidth}
\centerline{\includegraphics[width=0.99\linewidth]{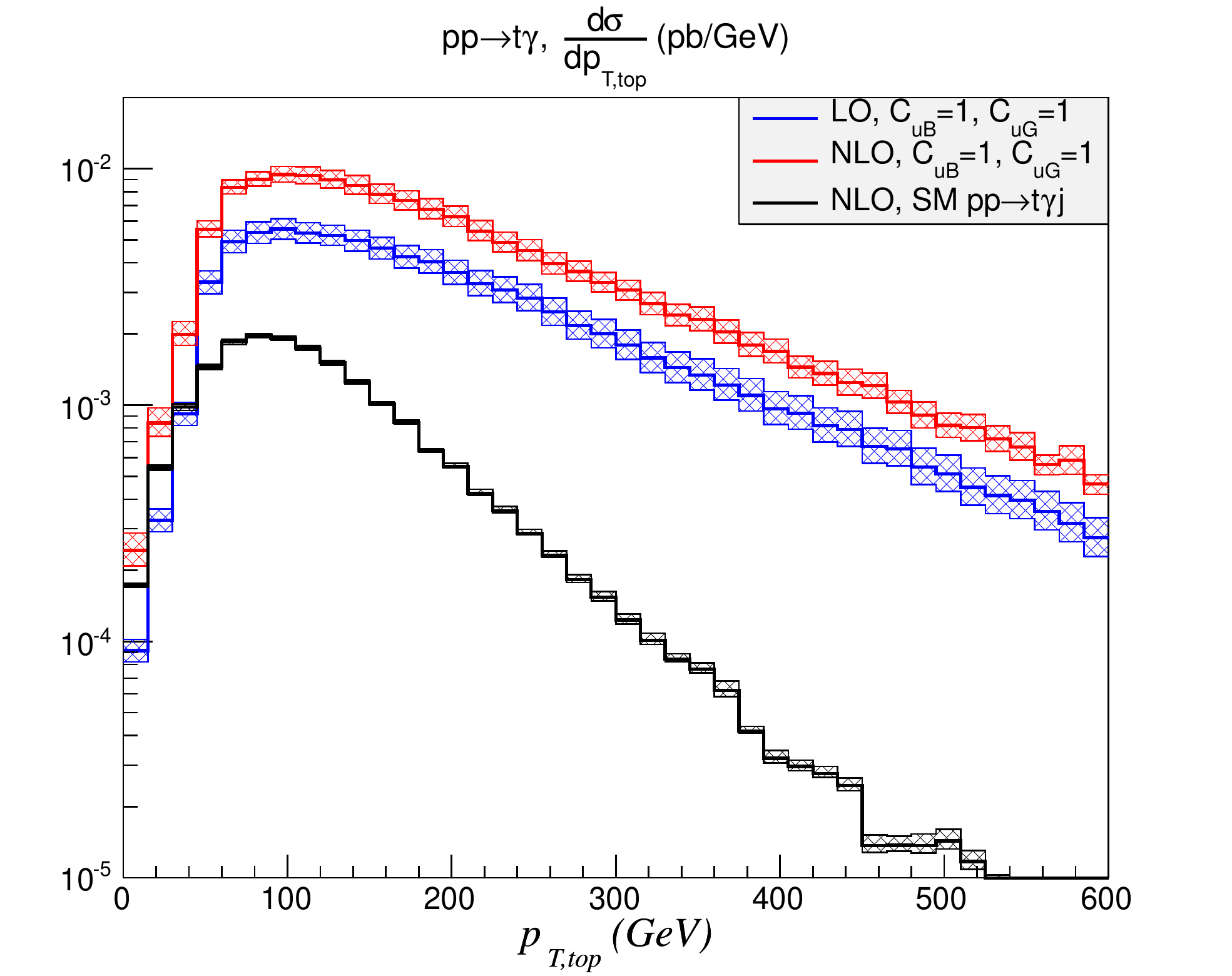}}
\end{minipage}
\hfill
\begin{minipage}{0.46\linewidth}
\centerline{\includegraphics[width=0.99\linewidth]{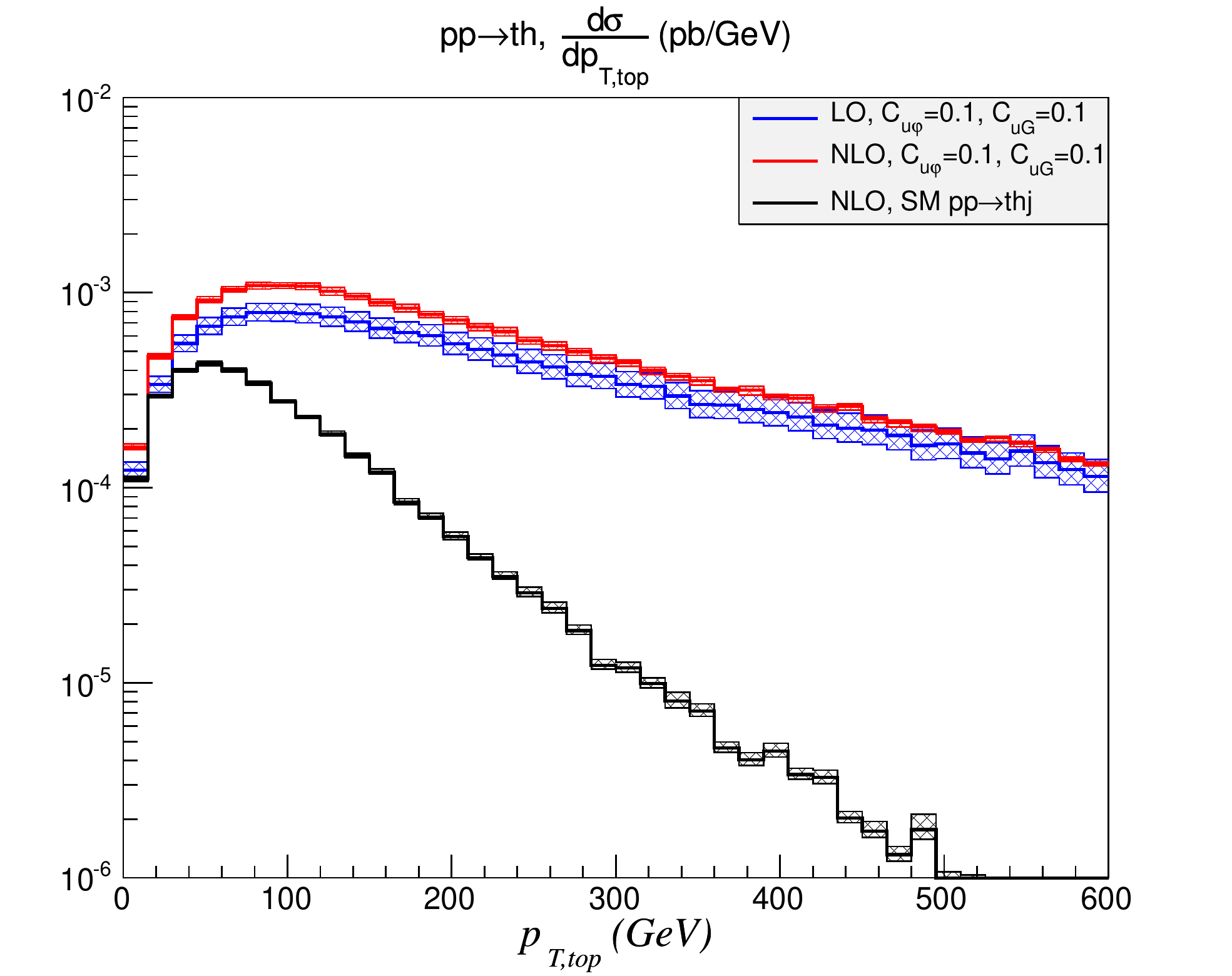}}
\end{minipage}
\caption{\label{fig:pptah}The $p_T$ distribution of the top quark in $pp\to
t\gamma$ (left), and in $pp\to th$ (right).  The SM background is from $t\gamma
j$ (left) and $thj$ (right) production.}
\end{figure}
Theoretical predictions for $t\gamma$, $tZ$ and $th$ production processes have
been computed in Refs.~\cite{Zhang:2011gh,Li:2011ek,Wang:2012gp} at NLO in QCD.
The NLO corrections are found to be at $\sim40\%$ to $80\%$ level.  We have
implemented the two-quark flavor-changing operators to the MG5\_aMC@NLO framework
\cite{Alwall:2014hca}, using FeynRules \cite{Alloul:2013bka}
and NLOCT \cite{Degrande:2014vpa}.
As a result, processes such as $pp\to tX$ where $X=j,\gamma,Z,h$ can be computed
automatically at NLO in QCD and are matched to parton shower simulation.
For illustration, we present in Fig.~\ref{fig:pptah} the $p_T$ distribution of
the top in $pp\to t\gamma$ and in $pp\to th$ at NLO.  
Because the calculation is automatic, a rich set of processes can be studied at
NLO with parton shower, including single top production in not only $pp$
collision but also $e^+e^-$ collision, and top-quark decay in single top and
$t\bar t$ production modes.  In future four-fermion operators involving the top
quark will also be available, allowing for more interesting processes to be
studied.  All these studies can give useful information about top-quark FCNC
couplings.

\section{Summary}

The top quark physics has entered a precision era.
Due to the scale separation between $\Lambda_{NP}$
and $m_t$, the EFT approach is a natural tool to study the indirect effects
from new physics.  In this talk we have discussed theoretical results for
top quark processes in the EFT framework.  We have focused on top-quark decay
and searches for top quark FCNC interactions, and discussed the need for a
global analysis for top-quark properties based on an EFT approach.  Moreover,
NLO predictions are now available for these processes, and in particular, the top-quark
FCNC operators are implemented to the MG5\_aMC@NLO framework.  These results
provide information and tools needed to perform a global analysis.

\ack
I would like to thank my collaborators, Celine Degrande, Gauthier Durieux, Fabio
Maltoni and Jian Wang.  This work is supported by the IISN ``Fundamental
interactions'' convention 4.4517.08.

\appendix
\section{Top-Quark FCNC Operators}
\label{sec:app}
Top FCNC interactions are characterized by the following two-quark operators:
\vspace{-4pt}
\newcommand{\FDF}{\left(\varphi^\dagger\overleftrightarrow{D}_\mu\varphi\right)}
\newcommand{\FDFI}{\left(\varphi^\dagger\overleftrightarrow{D}^I_\mu\varphi\right)}
\begin{equation}
  \small
  \begin{array}{lll}
    O_{\varphi q}^{(3,i+3)}=i\FDFI(\bar{q}_i\gamma^\mu\tau^IQ)
    &
    O_{\varphi q}^{(1,i+3)}=i\FDF(\bar{q}_i\gamma^\mu Q) 
    &
    O_{\varphi u}^{(i+3)}=i\FDF(\bar{u}_i\gamma^\mu t)
    \\
    O_{uB}^{(i3)}=g_Y(\bar{q}_i\sigma^{\mu\nu}t)\tilde{\varphi}B_{\mu\nu}
    &
    O_{uW}^{(i3)}=g_W(\bar{q}_i\sigma^{\mu\nu}\tau^It)\tilde{\varphi}W^I_{\mu\nu}
    &
    O_{uG}^{(i3)}=g_s(\bar{q}_i\sigma^{\mu\nu}T^At)\tilde{\varphi}G^A_{\mu\nu}
    \\
    O_{u\varphi}^{(i3)}=(\varphi^\dagger\varphi)(\bar{q}_it)\tilde\varphi
    \nonumber
  \end{array}
\vspace{-4pt}
\end{equation}
where the subscript $i$ is the generation of the quark field. $Q$ is the
third-generation doublet. For operators with $(i3)$ superscript, a similar set
of operators with $(3i)$ flavor structure can be obtained by interchanging
$(i3)\leftrightarrow (3i),\ t\leftrightarrow u_i$ and $Q\leftrightarrow q_i$.
$O_{\varphi q}^{(-,i+3)}$ is defined as $(O_{\varphi q}^{(1,1+3)}\!-\!O_{\varphi
q}^{(3,1+3)})/2$.  A full list including four-fermion operators
can be found elsewhere \cite{Zhang:2014rja}.


\section*{References}
\begin{thebibliography}{9}
  \bibitem{Grzadkowski:2010es} 
    Grzadkowski B, Iskrzynski M, Misiak M and Rosiek J 2010
    {\it JHEP} {\bf 1010} 085

  \bibitem{Jenkins:2013zja} 
    Jenkins E E, Manohar A V and Trott M 2013
    {\it JHEP} {\bf 1310} 087

  \bibitem{Jenkins:2013wua} 
    Jenkins E E, Manohar A V and Trott M 2014
    {\it JHEP} {\bf 1401} 035

  \bibitem{Alonso:2013hga} 
    Alonso R, Jenkins E E, Manohar A V and Trott M 2014
    {\it JHEP} {\bf 1404} 159

  \bibitem{Alwall:2014hca} 
    Alwall J, Frederix R, Frixione S, Hirschi V, Maltoni F, Mattelaer O, Shao H -S, Stelzer T, Torrielli P and Zaro M 2014
    {\it JHEP} {\bf 1407} 079

  \bibitem{Drobnak:2010ej} 
    Drobnak J, Fajfer S and Kamenik J F 2010
    {\it Phys. Rev.} D {\bf 82} 114008

  \bibitem{Zhang:2014rja} 
    Zhang C 2014
    {\it Phys. Rev.} D {\bf 90} 014008

  \bibitem{Chatrchyan:2013jna} 
    CMS Collaboration 2013
    {\it JHEP} {\bf 1310} 167

  \bibitem{Drobnak:2010wh} 
    Drobnak J, Fajfer S and Kamenik J F 2010
    {\it Phys. Rev. Lett.} {\bf 104} 252001

  \bibitem{Zhang:2010bm} 
    Zhang J J, Li C S, Gao J, Zhu H X, Yuan C P and Yuan T C 2010
    {\it Phys. Rev.} D {\bf 82} 073005

\bibitem{Chatrchyan:2013nwa} 
  CMS Collaboration 2014
  {\it Phys. Rev. Lett.} {\bf 112} 171802

  \bibitem{CMS:2014qxa} 
    CMS Collaboration 2014
    Combined multilepton and diphoton limit on t to cH
    {\it CMS-PAS-HIG-13-034}

  \bibitem{Greljo:2014dka} 
    Greljo A, Kamenik J F and Kopp J 2014
    {\it JHEP} {\bf 1407} 046

  \bibitem{Aad:2014dya} 
      ATLAS Collaboration 2014
      {\it JHEP} {\bf 1406} 008

  \bibitem{TheATLAScollaboration:2013vha} 
    ATLAS Collaboration 2013
    Search for single top-quark production via FCNC in strong interaction in $\sqrt{s}=8\,\,\mathrm{TeV}$ ATLAS data
    {\it ATLAS-CONF-2013-063}

  \bibitem{CMS:2014hwa} 
    CMS Collaboration 2014
    Search for anomalous single top quark production in association with a photon
    {\it CMS-PAS-TOP-14-003}

  \bibitem{CMS:2013nea} 
    CMS Collaboration 2013
    Search for Flavour Changing Neutral Currents in single top events
    {\it CMS-PAS-TOP-12-021}

  \bibitem{Zhang:2011gh} 
    Zhang Y, Li B H, Li C S, Gao J and Zhu H X 2011
    {\it Phys. Rev.} D {\bf 83} 094003

  \bibitem{Li:2011ek} 
    Li B H, Zhang Y, Li C S, Gao J and Zhu H X 2011
    {\it Phys. Rev.} D {\bf 83} 114049

  \bibitem{Wang:2012gp} 
    Wang Y, Huang F P, Li C S, Li B H, Shao D Y and Wang J 2012
    {\it Phys. Rev.} D {\bf 86} 094014
  
\bibitem{Alloul:2013bka} 
  Alloul A, Christensen N D, Degrande C, Duhr C and Fuks B 2014
  {\it Comput. Phys. Commun.}  {\bf 185} 2250

\bibitem{Degrande:2014vpa} 
  C.~Degrande 2014
  Automatic evaluation of UV and R2 terms for beyond the Standard Model Lagrangians: a proof-of-principle
  {\it Preprint 1406.3030 [hep-ph]}

\end{thebibliography}
\end{document}